\documentclass[aps,twocolumn]{revtex4-1}
\usepackage{bbold}
\usepackage{graphicx}
\usepackage{grffile}
\usepackage{dcolumn}
\usepackage{bm}
\usepackage{amsmath}
\usepackage{hyperref}
\hypersetup{colorlinks=true, urlcolor=blue}

\usepackage{hyperref}
\usepackage{epsfig}
\usepackage{newlfont}
\usepackage{amssymb}
\usepackage{amsfonts}
\usepackage{amsmath}
\usepackage{bm}

\def \be{\begin{equation}}
\def \ee{ \end{equation} }

\begin{document}
\renewcommand*{\DefineNamedColor}[4]{%
  \textcolor[named]{#2}{\rule{7mm}{7mm}}\quad
  \texttt{#2}\strut\\}

\definecolor{red}{rgb}{1,0,0}

\title{Tuning interaction strength leads to ergodic-nonergodic transition of quantum correlations in anisotropic  Heisenberg spin model}

\author{Utkarsh Mishra, R. Prabhu, Aditi Sen(De), and Ujjwal Sen}

\affiliation{Harish-Chandra Research Institute, Chhatnag Road, Jhunsi, Allahabad 211 019, India}

\begin{abstract}
We investigate the time dynamics of quantum correlations of the anisotropic Heisenberg model in a time-dependent magnetic field, in one-dimensional, ladder, and two-dimensional lattices.  We find that quantum correlation measures in the entanglement-separability paradigm are ergodic in these systems irrespective of system parameters.  However, information-theoretic quantum correlation measures can also be nonergodic, and exhibit a transition from nonergodic to ergodic behavior with 
the change of interaction strength in the direction of the magnetic field. 
We also observe that the transition point changes drastically as we go from one-dimensional and ladder lattices to the two-dimensional one.
\end{abstract}

\maketitle
\section{Introduction}

Statistical mechanical models provide a quantitative way to understand physical phenomena involving a large number of particles which are interacting among themselves. These models have been established as promising substrates in different physical systems for implementing many quantum information protocols which include, for example, one-way quantum computation \cite{RB} and  quantum communication tasks \cite{SBOSE}. The characterization, quantification and realization of quantum correlations in many-body systems are some of the main challenges in quantum information \cite{HHHH1,kmodi,FAZIORMP,ASDADP}. 

Quantum correlation concepts in multiparty systems can broadly be classified into two categories
-- entanglement-separability paradigm measures and information-theoretic ones. Quantum correlations of the first kind are established to be  useful resources for many quantum information tasks which include quantum dense coding \cite{BennettWiesner}, quantum teleportation \cite{Bennettetal}, and  secure quantum cryptography \cite{GRTZ}. 
Recently however, several non-classical phenomena have been discovered in which entanglement is absent \cite{nlwe,KnillLaflamme,Animesh,others,MLSM}. To understand and quantify the resource necessary for exhibiting such non-classicality, information-theoretic quantum correlation measures like quantum discord \cite{OllWH,HV} and quantum work-deficit \cite{OHHH} have been proposed. 

Measures of both the paradigms have proven to be  advantageous in investigations of cooperative  physical phenomena observed in many-body systems \cite{HHHH1,kmodi,ASDADP,FAZIORMP,DILL}. Due to the paucity of analytical as well as numerical methods to solve quantum spin models, most of these considerations are restricted to the ground state or the thermal state of the system. While it is important to understand the quantum correlation properties of these ``static states" of the system, the time-evolution of the system is an integral part of several quantum information processing protocols, a prominent example being the one-way quantum computer \cite{RB}.

Properties like magnetization, susceptibility, classical and quantum correlations in the static states of the isotropic Heisenberg model have been studied extensively, both theoretically and experimentally \cite{FAZIORMP,ASDADP,Lemmensa03, MONS_RMP}. The model can be exactly solved by the Bethe ansatz \cite{Bethe}. Variation of different physical parameters in this model leads to the appearance of rich phases \cite{lhuillier05,Fisher06}, like spin-liquid, resonating valence bond states, etc. Moreover, such models can now be created in the laboratories in a controlled way by using e.g., photons \cite{XYZphoton}, trapped ions \cite{XYZtrapion}, and cavity QED \cite{XYZcQED}. However, numerical simulations or approximate methods are the only techniques that can be used to investigate properties of the time-evolved states of this model. Here we investigate the behavior of quantum correlations of the evolved state as well as the equilibrium state in the anisotropic Heisenberg model in low-dimensional systems, under the influence of time-dependent magnetic fields and temperature. In particular, we observe collapse and revival of quantum correlation measures of the evolved state in this system. 

The usual statistical mechanical description of a physical quantity is valid only when the time-average of the quantity matches with its ensemble average, and in that case, the physical quantity is termed as ergodic. Ergodicity of physical quantities in spin models has been of interest to researchers for a long time  \cite{Perk,Mazur,BMD,AUL,Kais}. In particular, the question of ergodicity of physical quantities like magnetization, classical correlations, and entanglement in quantum XY spin chains have been investigated \cite{Mazur,BMD,Perk,AUL,prabhu}.

Here we consider the validity of the statistical mechanical description of quantum correlation measures of anisotropic Heisenberg models in one-dimension (\(1D\)), ladder and two-dimension (\(2D\)). 
Specifically, we find that the entanglement measures remain ergodic, irrespective of the initial strength of the applied magnetic field in the $z$-direction and the interaction strengths, whereas for intermediate values of the initial magnetic field, the information-theoretic measures like quantum discord and quantum work-deficit show a transition from nonergodic to ergodic behavior, with the tuning of the strength of the two-body interaction in the $z$-direction. The results hold irrespective of the relative strength (``anisotopy") of the $xx$- and $yy$-interactions. However, the transition point depends on the $xy$-anisotropy (i.e., the parameter that controls the relative strength of the $xx$- and $yy$-interactions) and the strength of the magnetic field.  

The paper is organized as follows. In Sec. II, we give a brief description of the model under investigation. In Sec. III, we discuss about the canonical equilibrium state, the time-evolved state, and ergodicity. Calculation of  bipartite quantum correlations requires the two-site density matrix of the system.
We discuss properties of single-site and two-site density matrices for the equilibrium as well as the time-evolved state in Sec. IV. Measures of quantum correlations for both the paradigms are defined in Sec. V. 
We present our results  in Sec. VI (for $1D$), in  Sec. VII (for ladder), and in Sec. VIII (for $2D$). We conclude in Sec. IX.
 
\section{The Model}

We consider a system of $N$ quantum spin-$\frac{1}{2}$ particles arranged in a lattice with unequal nearest-neighbor interactions along $x,\, y,\, \mbox{and}\, z$ directions. It is therefore the antiferromagnetic anisotropic Heisenberg model or the XYZ model, and is given by 
\begin{eqnarray}
H_{{int}}=\frac{1}{4}\sum [J_{x}\sigma_{\vec{i}}^{x}\sigma_{\vec{j}}^{x}+J_{y}\sigma_{\vec{i}}^{y}\sigma_{\vec{j}}^{y}+J_{z}\sigma_{\vec{i}}^{z}\sigma_{\vec{j}}^{z}],
\label{eq:Hamil}
\end{eqnarray}
where $\sigma^a_{\vec{i}}\, (a=x,y,z)$ are the Pauli spin matrices at the site $\vec{i}$ of the spin lattice, and $J_x,\, J_y,\, \mbox{and}\, J_z$ represent the coupling constants in the $x,\, y \, \mbox{and}\, z$ directions respectively. 
The summation in Eq. (\ref{eq:Hamil}) runs over all nearest-neighbor pairs on the lattice. Periodic boundary conditions are assumed in all cases considered in this paper. We will consider systems of quantum spins arranged in lattices in different low dimensions. Assigning different relations among $J_x,\, J_y,\, \mbox{and}\, J_z,$ in the above Hamiltonian lead to various other well-known spin models, including the isotropic Heisenberg model for which  $J_x=J_y=J_z$, and the anisotropic XY model for which $J_x \neq J_y, \, J_z=0$. To check for ergodic properties of different physical quantities of these Heisenberg spin models, we will consider the initial state of the evolution to be the canonical equilibrium state at the initial instant (see discussion in the succeeding section). A non-trivial evolution of the system can be obtained in this case by introducing 
a magnetic field represented by $H_{{mag}}$, in such a way that $[H_{{int}},H_{{mag}}]\neq 0$. Hence the total Hamiltonian can now be written as
\begin{equation}
 H(t)=H_{{int}}-h(t)H_{{mag}}.
\end{equation}
For the present paper, we choose $J_x=J(1+\gamma)J,\, J_y=J(1-\gamma),\, J_z=J\delta$ and \(H_{mag}=\frac{J}{2}\sum\sigma_{\vec{i}}^{z}\), with the summation running over all sites of the lattice. Here \(J>0\) is assumed to have the dimension of energy, while \(\gamma\) and \(\delta\) are dimensionless system parameters. Here $\gamma$ represent $xy$-anisotropy. For brevity, we will sometimes call it simple as ``anisotropy". The time-dependence of the applied magnetic field is of the form 
\begin{equation}
 h(t)= \left\{
 \begin{array}{cc}
 a, & t\leq 0  \\
 0, & t>0,
\end{array}\right.
\end{equation}
where $a \neq 0 $ is a dimensionless parameter. Here \(t\) represents the time. Therefore, the total Hamiltonian is given by
\begin{eqnarray}
 H(J,\gamma,\delta,h(t))&=\frac{J}{4}\sum[(1+\gamma)\sigma_{\vec{i}}^{x}\sigma_{\vec{j}}^{x}+(1-\gamma)\sigma_{\vec{i}}^{y}\sigma_{\vec{j}}^{y}\nonumber\\
 &+\delta\sigma_{\vec{i}}^{z}\sigma_{\vec{j}}^{z}] - \frac{J}{2}h(t)\sum\sigma_{\vec{i}}^{z}.
\label{Hamilwithfield}
\end{eqnarray}
When $h(t)=0$ and $ J_x= J_y = J_z,$ the above Hamiltonian is exactly solvable by using Bethe ansatz \cite{Bethe} by which the ground state energy can be obtained \cite{Yosida}. However, there exists no such exact solution for the anisotropic Heisenberg model. Moreover, we wish to study the evolution of the system and hence require the single site- and two-site properties of the entire energy spectrum of the system at a given time. Hence, to study the statistical mechanical properties of such systems at finite temperature,
we opt for exact diagonalization using numerical simulations.

\section{Statistical Mechanical Properties}

In this paper, we aim to study the statistical mechanical properties of the anisotropic Heisenberg model in time-dependent external magnetic fields. The statistical mechanical notions like canonical equilibrium state, time-evolved state and ergodicity will be briefly defined in this section, mainly to set the terminology and the notations. In particular, we introduce a quantity called the ``ergodicity score" which helps us to quantify the degree to which a physical quantity is possibly nonergodic.

\subsection{Time-evolution}

For the quantum spin system, described by the Hamiltonian in Eq. (\ref{Hamilwithfield}), we denote the canonical equilibrium state of the system, at time \(t\), as \(\rho_{eq}^{\beta}\), and is given by
\begin{equation}
\rho_{eq}^{\beta}(t) = \frac{\mbox{exp}(- \beta H(t))}{\cal Z},
\label{eq:equilib}
\end{equation}
where \(\cal Z\) is the partition function, 
\[\cal Z = \mbox{tr}[\mbox{exp}(-\beta H(t))], \]
and \(\beta = \frac{1}{k_{B}T}\), with \(k_B\) being the Boltzmann constant. \(T\) represents the absolute temperature.


The canonical equilibrium state can evolve due to the application of external ``disturbances",
like switching on of the magnetic field across the system.  In our case, the evolution of the system is governed by the Hamiltonian given in  Eq. (\ref{Hamilwithfield}). We assume that the system is in contact with a heat bath at temperature \(T'\) for a long time until \(t=0\). We assume that the contact is in the canonical sense, so that the system and the heat bath exchange energy (under the normal average energy constraint), but do not exchange particles. We assume that this contact leads the system to the canonical equilibrium state at \(t=0\), i.e., the state of the system at \(t=0\) is \(\rho_{eq}^{\alpha}(0)\), where \(\alpha=\frac{1}{k_{B}T'}\). For \(t>0\), the magnetic field is switched off, and we consider the situation where the contact with the heat bath is also cut off for all times \(t>0\). The system therefore starts evolving according to the Schr\"{o}dinger equation governed by the Hamiltonian in Eq. (\ref{Hamilwithfield}), with the initial state of this evolution being \(\rho_{eq}^{\alpha}(0)\), and we denote the corresponding evolved state as $\rho^{\alpha}(t)$. Note here that $\rho^{\alpha}_{eq}(t=0)=\rho^{\alpha}(0)$.

\subsection{Ergodicity and Ergodicity Score}

To check whether a given physical quantity \(\cal  Q\) is ergodic, we consider the value of \(\cal Q\) in the evolved state at a ``large time".
The time of evolution, \(t_{l}\), is termed as large, for the physical quantity \(\cal Q\), if 
(i) there are no fluctuations in the physical quantity $\cal Q$ with respect to time for \(t>t_{l}\), or if (ii) the fluctuation amplitude of $\cal Q$ with respect to time is smaller than the required precision level, for \(t>t_{l}\), or if (iii) the fluctuations of $\cal Q$ with respect to time is of a constant amplitude. We are interested in the time-average of the physical quantity ${\cal Q}$ at large-times. For the cases (i) and (ii), an explicit time-averaging is not required, as the system dynamics brings the quantity ${\cal Q}$ to its time-averaged value. For the case (iii), an explicit time-averaging for times \(t>t_{l}\) is required.
We now ask whether there exists a temperature $T'$, at which the large-time time-averaged value of a physical quantity $\cal Q$ in the evolved state is equal to the value of same physical quantity in the equilibrium state at temperature \(T\) at large-time. The physically relevant range of $T$ can be considered as up to an order of magnitude of the initial temperature $T'$. This difference between $T$ and $T'$ is, for example, to allow for possible errors in an experimental realization of the physical system or some typical theoretical effective standard deviation in that system.

If the time-average of a physical quantity is the same as the ensemble average, the quantity is said to be ergodic. 
Such a study is therefore based on the comparison of the large-time time-averaged value, \({\cal Q}^\infty (T', a)\), with the canonical equilibrium value, \({\cal Q}^{\scriptsize{can}}(T, h(t=\infty))\). Note that these quantities also depend on the system parameters $J,\, \gamma,\,\mbox{and}\, \delta$. The physical quantity ${\cal Q}$ is therefore said to be ergodic if 
\begin{equation}
\label{raat-duto-egaro}
{\cal Q}^\infty(T{'},a) = {\cal Q}^{\scriptsize{can}}(T,h(t=\infty)). 
\end{equation}
Otherwise, it is termed as nonergodic. 


Let us now introduce a quantity which can quantify the degree to which a given physical quantity, \(\cal Q\) fails to be ergodic. We call it the ``ergodicity score", and define it as 
\begin{equation}
{\cal Q}(\tilde \delta, \alpha) = \max[0,{\cal Q}^{\infty}(T',a)-\max_T{\cal Q}^{can}(T, h(t=\infty))]
\end{equation}
where $\tilde \delta$ denotes the aggregate of all physical parameters required to define the Hamiltonian of the system under consideration. For the system considered in this paper, $\tilde \delta$ consists of $J,\, h,\, \delta$ and $\gamma$. We remember that $\alpha=\frac{1}{k_B T'}$. The maximization over \(T\) is for all \(T\) that falls in the physically relevant range around \(T{'}\), as discussed earlier. Note therefore that a
non-zero value of  $\eta^{\cal Q}$ implies that ${\cal Q}$ is nonergodic and that a vanishing ${\cal Q}$ indicates ergodicity. 

\section{Single- and Two-site density matrices of time-dependent Heisenberg Model}

To analyze the ergodic properties of quantum correlations, let us now find the general form of the single- and two-site density matrices of equilibrium and evolved states of the Hamiltonian given in Eq. (\ref{Hamilwithfield}). The general single-site density matrix is given by
\begin{equation}
 \rho^{1}=\frac{1}{2} [\mathbb{I} + \vec{m} . \vec{\sigma}],
\end{equation}
where $\mathbb{I}$ is the $2\times 2$ identity matrix, and $\vec{ m}$ = \mbox{tr}[$\rho^{1} \vec{ \sigma}$] is the magnetization vector. If the entire system is of $N$ qubits, then the single-site density matrix can be obtained by tracing out $N-1$ parties. For a periodic lattice, tracing out of any $N-1$ qubits will lead to the same single-site density matrix. In the equilibrium state, since $\rho^{\beta *}_{eq}(t) = \rho^{\beta}_{eq}(t)$, where the complex conjugation has been taken in the computational basis, $m_{x}=0$. Moreover, in this case, $ m_{y}=0 $, since $[H,\prod_{i}\sigma^{z}_{i}]= 0 $. Therefore, the single-site density matrix for the equilibrium state reduces to
\begin{equation}
 \rho^{1}_{eq}(t)=\frac{1}{2}[\mathbb{I}+m_{z}^{eq}(t)\sigma^{z}].
\end{equation}
where we have hidden the dependence on temperature in the notation. The single-site density matrix for the evolved state also turns out to be $\rho^{1}(t)=\frac{1}{2}( \mathbb{I} + m_{z}(t) \sigma^{z}) $, using the Wick's theorem \cite {BMD,LSM}.

The nearest-neighbor two-site density matrix can be written, in general, as 
$$\rho^{12}=\frac{1}{4}[\mathbb{I}\otimes \mathbb{I}+ \vec{m}.\vec{\sigma}\otimes \mathbb{I} + \mathbb{I}\otimes  \vec{m}.\vec{\sigma} + \sum_{i,j=x,y,z} T^{i j}(\sigma^{i} \otimes \sigma^{j})]$$
where $T^{ij} = \mbox{tr}[(\sigma^{i} \otimes \sigma^{j}) \rho^{12}]$ represent the two-site correlation functions. Since periodic boundary conditions are assumed, the nearest-neighbor state \(\rho^{12}\) is independent of which two neighboring sites are chosen for constructing the nearest-neighbor state. Due to the form of the single-site density matrices that has already been derived, the two-site density matrices for both equilibrium and evolved states reduces to
\begin{equation*}
 \rho^{12}=\frac{1}{4}[\mathbb{I}\otimes \mathbb{I}+ m_{z}(\sigma^{z}\otimes \mathbb{I} + \mathbb{I} \otimes \sigma^{z})+\sum_{i,j=x,y,z}T^{i j}(\sigma^{i} \otimes \sigma^{j})].
\end{equation*}
Using Wick's theorem, we can show that  all off-diagonal correlations 
vanish  for the equilibrium state. However, for the evolved state, only $xz$- and $yz$-ones vanish.

\section{Measures Of  quantum correlation}

We will now quickly define the measures of quantum correlations used in this paper. We will introduce two measures within the entanglement-separability paradigm, namely logarithmic negativity and concurrence. We will subsequently define two information-theoretic quantum correlation measures, viz. quantum discord and quantum work-deficit. 

\subsection{Logarithmic Negativity}

Given a bipartite quantum state, $\rho^{AB}$, shared between two parties $A$ and $B$, the logarithmic negativity \cite{vedWer} quantifies the amount  of entanglement present in the bipartite state. The definition of logarithmic negativity is based on negativity, $N(\rho^{AB})$, which is defined as the sum of the absolute values of the negative eigenvalues of the partial transposed density matrix \cite{peres} of the bipartite state $\rho^{AB}$.
The logarithmic negativity (LN) is defined as
\begin{equation}
 E_{N}(\rho^{AB})=\mbox{log}_{2}[2 N(\rho^{AB})+1].
\end{equation}
For two qubit states, LN is positive if and only if the state is entangled \cite{peres}.

\subsection{Concurrence}

For two-qubit states, concurrence is another useful entanglement measure \cite{HW}. For a two-qubit mixed bipartite state $\rho^{AB}$, it is defined as 
\begin{equation}
 C(\rho^{AB})= \max [0,\lambda_{1}-\lambda_{3}-\lambda_{3}-\lambda_{4}],
\end{equation}
where $\lambda_{1},\lambda_{2},\lambda_{3},\lambda_{4}$ are the  square roots of the eigenvalues of $\rho^{AB}\tilde \rho^{AB}$ in decreasing order and 
$\tilde\rho^{AB} = [\sigma^{y}\otimes \sigma^{y})\rho^{* AB }(\sigma^{y}\otimes \sigma^{y}]$, with the complex conjugation being taken in the computational basis. The maximum is taken to ensure that concurrence is zero for separable states. The measure is non-zero for all entangled states. 

\subsection{Quantum Discord}

Quantum discord \cite{OllWH,HV} is an information-theoretic measure of quantum correlation. It is defined as
 \begin{equation}
{\cal D}(\rho^{AB})=I(\rho^{AB}) - J(\rho^{AB}).
\end{equation}
Here $I(\rho^{AB})$ and $J(\rho^{AB})$ are equivalent in classical information theory, where they both represent the mutual information between two random variables.  In the quantum world, the first term represents the total correlation of the bipartite state $\rho^{AB}$, and is given by 
$$I(\rho^{AB})=S(\rho^{A})+S(\rho^{B})-S(\rho^{AB}),$$ 
where $S(\rho) = - \mbox{tr} [\rho \log_2 \rho]$ is the von Neumann entropy of a quantum state $\rho$, and $\rho^{A}$ and $\rho^{B}$ are the reduced density matrices of $\rho^{AB}$. The second term, $J(\rho^{AB})$ in the definition of quantum discord can be argued as the amount of classical correlation present in $\rho^{AB}$, and is defined by
$$J(\rho^{AB})=S(\rho^{A}) - S(\rho^{A|B}).$$
Here $ S(\rho^{A|B}) =  \min_{\{B_{i}\}} \sum_{i}p_{i}S(\rho^{A|i})$ is the conditional entropy of $\rho^{AB}$, when the rank-\(1\) projection-valued measurement, \(\{B_{i}\}\), is performed on the $B$-part of the system, with $\rho^{A|i}=\mbox{tr}_{B}[(\mathbb{I}^{A}\otimes B_{i})\rho^{AB}(\mathbb{I}^{A}\otimes B_{i})]$, $p_{i}=\mbox{tr}_{AB}[(\mathbb{I}^{A}\otimes B_{i})\rho(\mathbb{I}^{A}\otimes B_{i})]$, and with $\mathbb{I}^A$ being the identity operator on the Hilbert space of $A$.

\subsection{Quantum Work-Deficit}

Another information-theoretic measure of quantum correlation is the quantum work-deficit, which is defined  as the difference between the amount of work extractable from a shared state by global and local quantum heat engines \cite{OHHH}. It is possible to quantify the amount of work that can be extracted from a bipartite state \(\rho^{AB}\) by global operations as
\begin{equation}
 I_{G}(\rho^{AB})=N-S(\rho^{AB}),
\end{equation}
where $N$ is the logarithm (base 2) of the dimension of the Hilbert space on which $\rho^{AB}$ is defined. 
It can be interpreted as the number of pure qubits that can be extracted from $\rho^{AB}$ by global operations on the state, and that consists of an arbitrary 
sequence of unitary and dephasing operations. Such operations are called ``closed global operations".
Let us now define ``closed local operations and classical communication (CLOCC)''. It  consists of local unitaries, local dephasing, 
and sending the dephased state from one party to other. The number of pure qubits  that can be extracted by CLOCC is given by,
\begin{equation}
 I_{L}(\rho^{AB})=N -\mbox{inf}_{\Lambda \epsilon \mbox{\scriptsize{CLOCC}}}[S({\rho{'}^{A}}) - S({\rho{'}^{B}})],
\end{equation}
where $S(\rho{'}^{A})= S(\mbox{tr}_{B}[\Lambda(\rho^{AB})])$ and $S(\rho{'}^{B})= S(\mbox{tr}_{A}[\Lambda(\rho^{AB})])$.
The quantum work-deficit is  defined as
\begin{equation}
 W_D(\rho^{AB})=I_{G}(\rho^{AB})-I_{L}(\rho^{AB}).
\end{equation}

In the next sections, our aim is to study the ergodicity of these quantum correlations in the anisotropic Heisenberg models of different lattice geometries. 
The lattices considered are the chain, the ladder, and the two-dimensional square lattice. Periodic boundary conditions is used in all cases.

\section{Quantum Heisenberg XYZ spin chain with magnetic field}

In this section, we investigate the statistical mechanical properties of quantum correlation measures in the one-dimensional quantum spin-\(\frac{1}{2}\) lattice described by the Hamiltonian in Eq. (\ref{Hamilwithfield}). The isotropic antiferromagnetic Heisenberg model in one-dimension provides an understanding of the spin-spin correlation functions and suppression of long range magnetic order in spin-liquids. Moreover, some materials like $\mbox{Sr}_{2}\mbox{CuO}_{3}\, \mbox{and}\, \mbox{SrCuO}_{2}$ mimic the Heisenberg spin chain \cite{Lemmensa03}. Recently developed techniques make it possible to realize this model in physical systems like photons \cite{XYZphoton}, trapped ions \cite{XYZtrapion}, and cavity QED \cite{XYZcQED}.
Entanglement in the ground and the thermal states of the Heisenberg model have been studied \cite{thermalentHei}.

\subsection{Quantum correlations in equilibrium and evolved states}

\begin{figure}
 \includegraphics[angle=270,width=4.2cm]{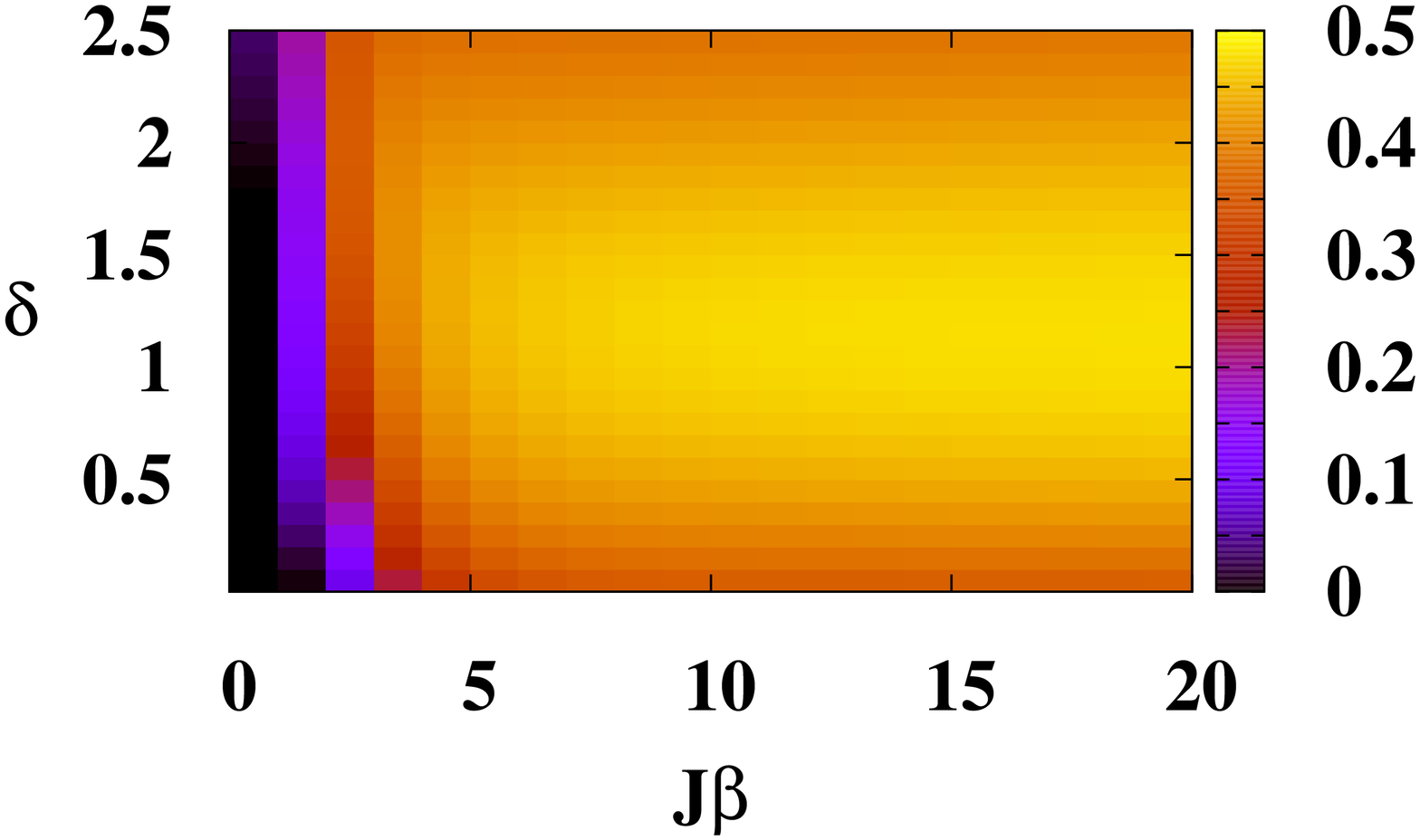}
 \includegraphics[angle=270,width=4.2cm]{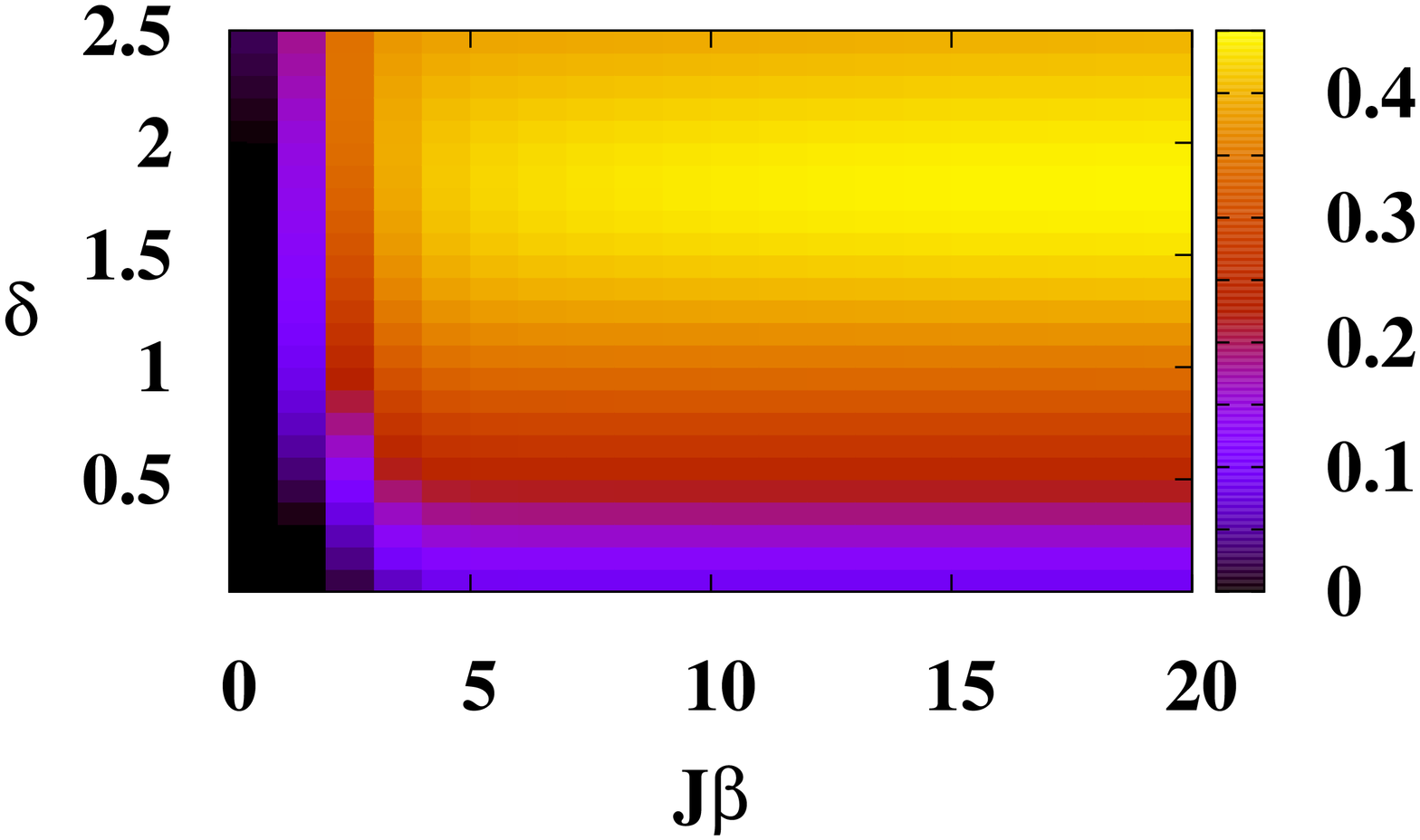}\\

 \includegraphics[angle=270,width=4.2cm]{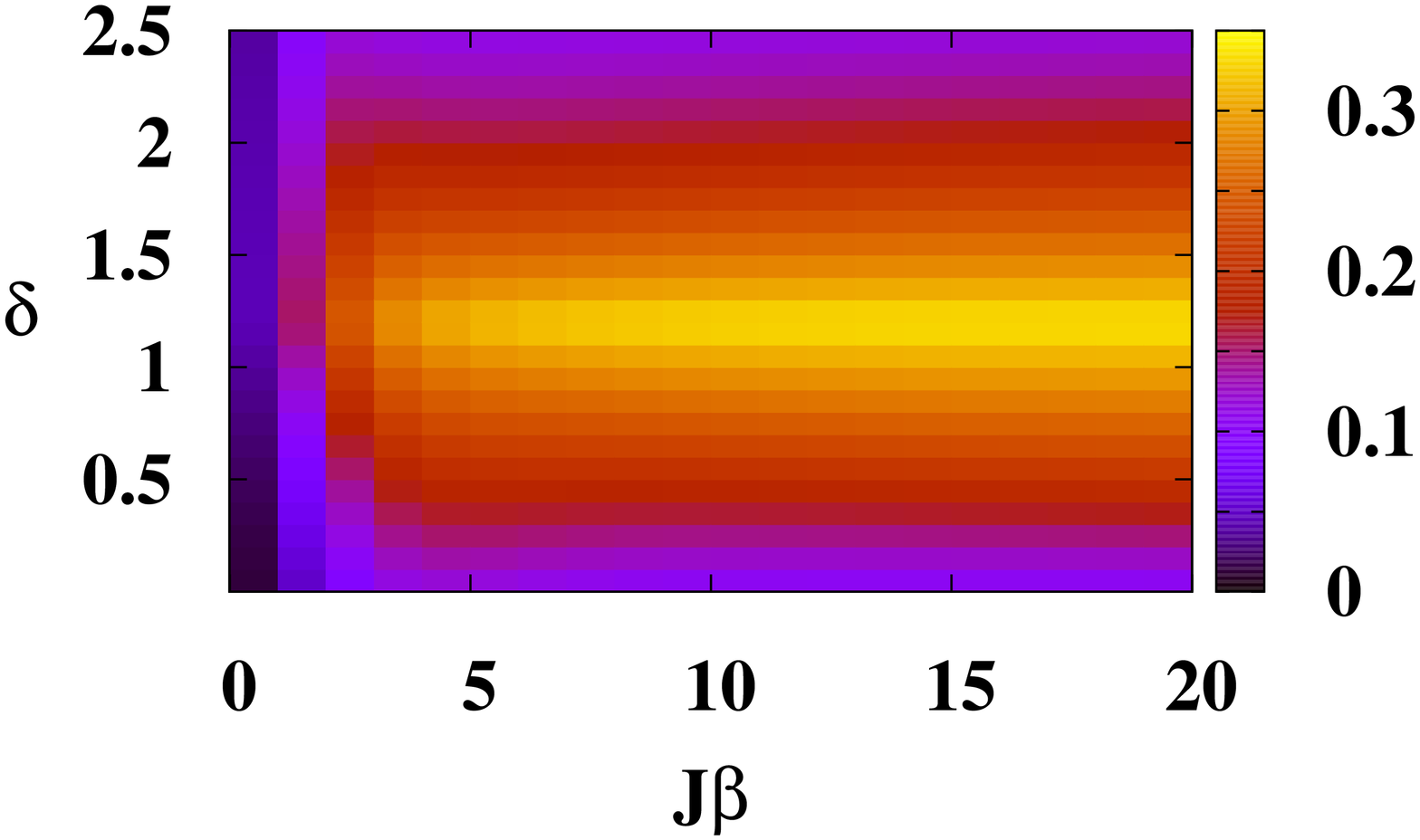}
 \includegraphics[angle=270,width=4.2cm]{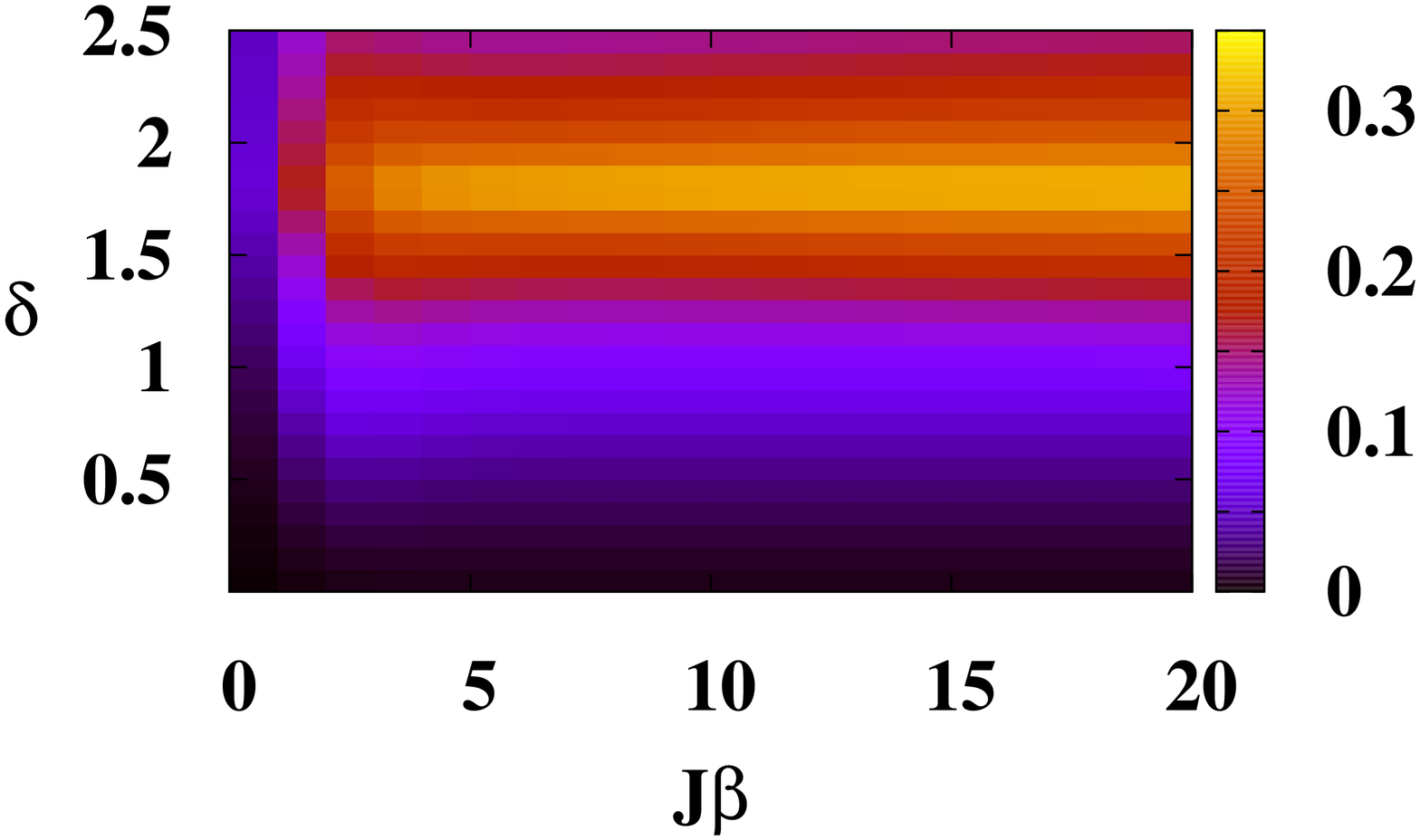} 
 \caption{(Color online) Behavior of quantum correlations in the equilibrium state. We plot quantum correlation measures of nearest-neighbor reduced states of the canonical equilibrium states, for a system of 12 quantum spin-$\frac{1}{2}$ particles arranged as a ring and described by the Hamiltonian $H$ with respect to  $J\beta$, and the relative strength of the $zz$-interaction, \(\delta\), for different values of \(\gamma\). The top plots are for logarithmic negativity and the bottom ones are for quantum discord. The left plots  are for \(\gamma=0.2\) while the right one are for \(\gamma=0.8\). Quantum discord is measured in bits. All other axes in the figures correspond to dimensionless parameters.
 }
 \label{fig:eqentgadel}
\end{figure}

For any system, that is in its canonical equilibrium state, all quantum correlations vanish when the temperature goes to infinity. Measures that are defined within the entanglement-separability paradigm typically vanish even for moderately high temperatures while information-theoretic measures like quantum discord  goes to zero asymptotically with the increase of temperature. This feature is retained by the system described by the Hamiltonian in Eq. (\ref{Hamilwithfield}), on an one-dimensional lattice with periodic boundary conditions. This shows that information-theoretic quantum correlation measures are more robust to temperature when compared to entanglement-separability measures. Moreover, we observe that
the entanglement of the nearest-neighbor reduced state of the canonical equilibrium state behaves differently with temperature in different ranges of $\gamma$ and $\delta$. See Figs. \ref{fig:eqentgadel}(a) and \ref{fig:eqentgadel}(b). In particular, we find that for fixed low values of the anisotropy, $\gamma$, the  entanglement saturates to a value with increasing $\beta$, and this saturated value is more or less independent of $\delta$, the relative strength of the $zz$-interaction. 
However, when $\gamma$ is relatively high, entanglement saturates to a low value for small $\delta$, 
while for high $\delta$, it saturates to a higher value. On the other hand, quantum discord saturates to a low value with decreasing temperature 
for small $\delta$ as well as for high $\delta$, while it saturates to a high value for intermediate values of $\delta$ (see Figs. \ref{fig:eqentgadel}(c) and \ref{fig:eqentgadel}(d)). This behavior of quantum discord is true for all values of $\gamma$. However, 
with the increasing of the value of $\gamma$, the point where the maximum value of quantum discord is obtained, shifts to higher values of $\delta$. 
We have performed calculations also for concurrence and quantum work-deficit, and they have qualitatively  similar features as logarithmic negativity and quantum discord respectively.

\begin{figure*}
\centering
 \includegraphics[width=0.35\textwidth]{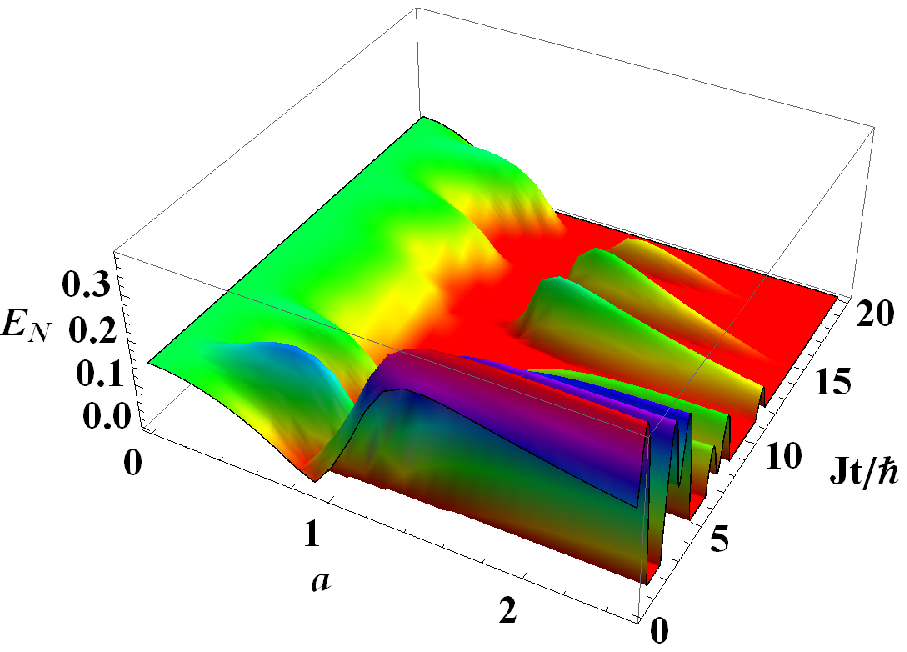}
 \includegraphics[width=0.35\textwidth]{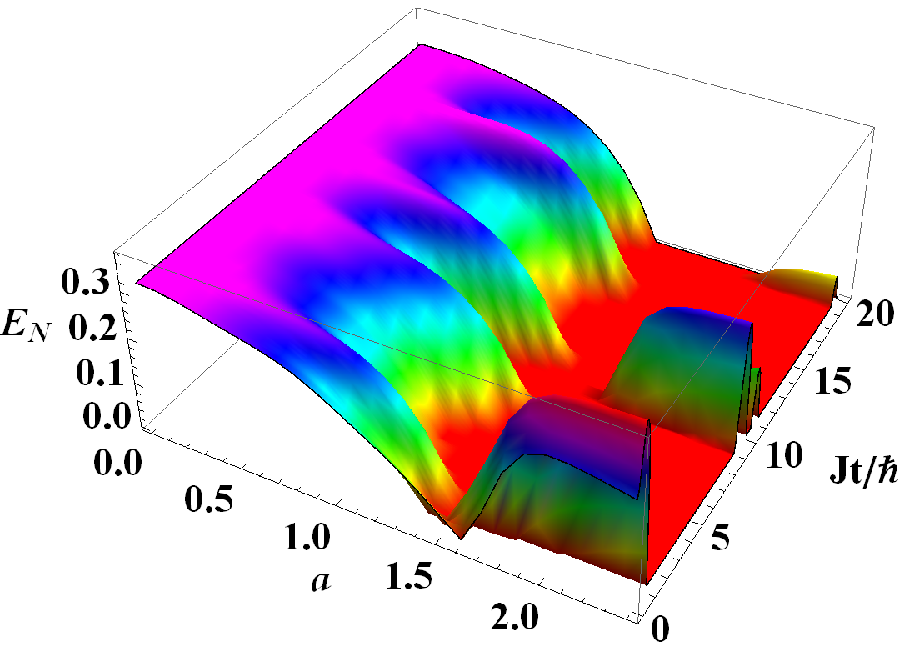}\\
 \includegraphics[width=0.35\textwidth]{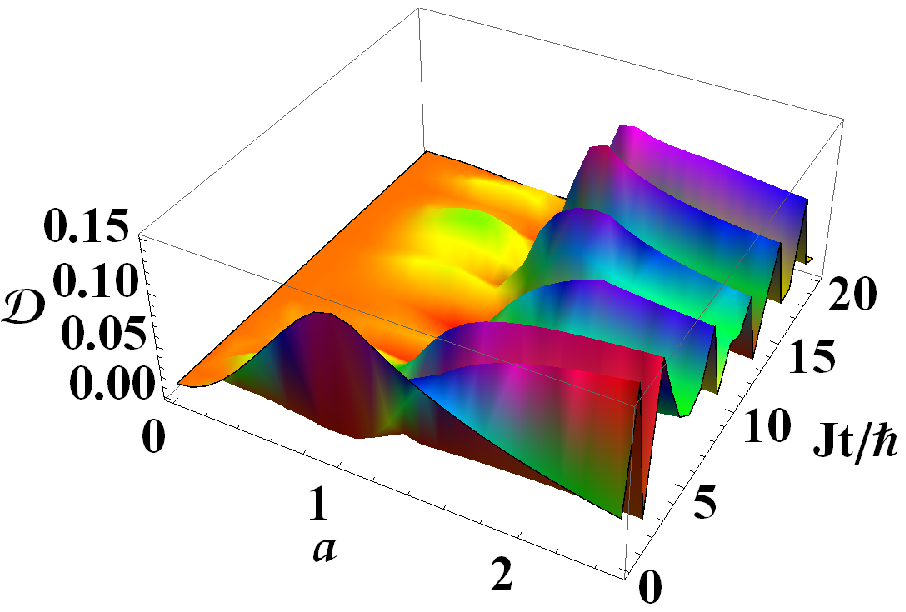}
 \includegraphics[width=0.35\textwidth]{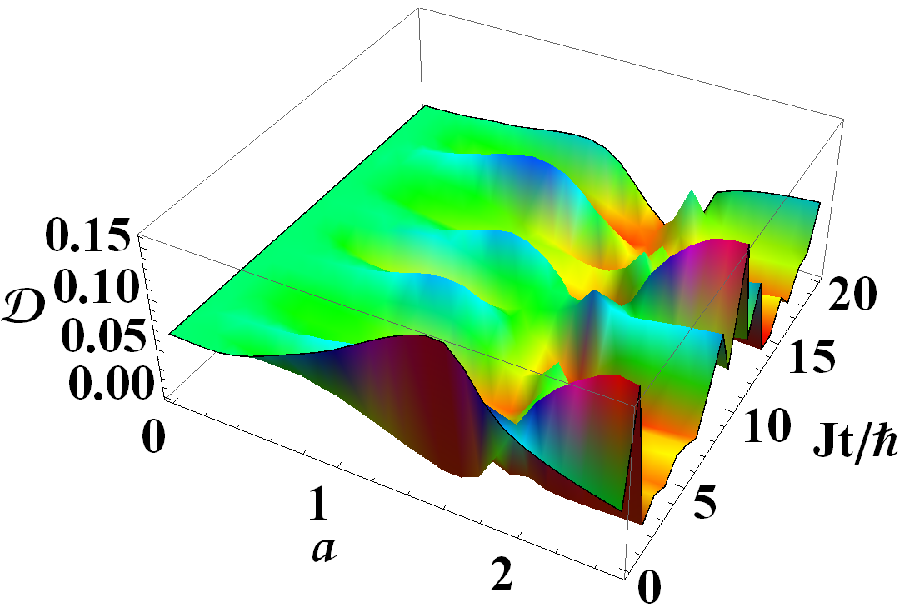}
 \caption{(Color online) Quantum correlations of the time-evolved states. The system under consideration is the same as in Fig. \ref{fig:eqentgadel}, but for 8 spins. The evolution is assumed to begin in the equilibrium state at $t=0$ and at an exemplary value of the temperature given by $J\alpha=20$. Logarithmic negativity (top plots) and quantum discord (bottom plots) of the nearest-neighbor reduced states of the time-evolved states, are plotted against the initial magnetic field, $a$, and $\frac{Jt}{\hbar}$, for different values of \(\delta\). Here we choose \(\gamma=0.8\).
 The left plots are for \(\delta=0.2\) and the right ones are for  \(\delta=0.8\). All axes correspond to dimensionless quantities except those for quantum discord, which is measured in bits.
 }
 \label{fig:evoldistimedelta}
\end{figure*}

Let us now discuss the time-dynamics of entanglement and other quantum correlations in the nearest-neighbor state. 
For the discussion, we choose $\gamma=0.8$. However, the behavior remains the same for other moderate values of $\gamma$.
The entanglement measures collapse and revive non-periodically with time, when $\delta$ is small. See Fig. \ref{fig:evoldistimedelta}(a), where we can view this feature for logarithmic negativity. 
For intermediate values of $\delta$, revival of entanglement occurs less frequently (Fig. \ref{fig:evoldistimedelta}(b)). 
For very high $\delta$, the model is ``Ising-like", and the entanglement as well as other quantum correlation measures collapse and revive periodically with time. 
The non-periodic collapse and revival behavior persists up to moderate values of $\delta$ for the information-theoretic quantum correlation measures like quantum discord. See Figs. \ref{fig:evoldistimedelta}(c) and \ref{fig:evoldistimedelta}(d).
 
\subsection{Statistical mechanical properties of quantum correlation measures}

\begin{figure}
 \includegraphics[width=0.45\textwidth]{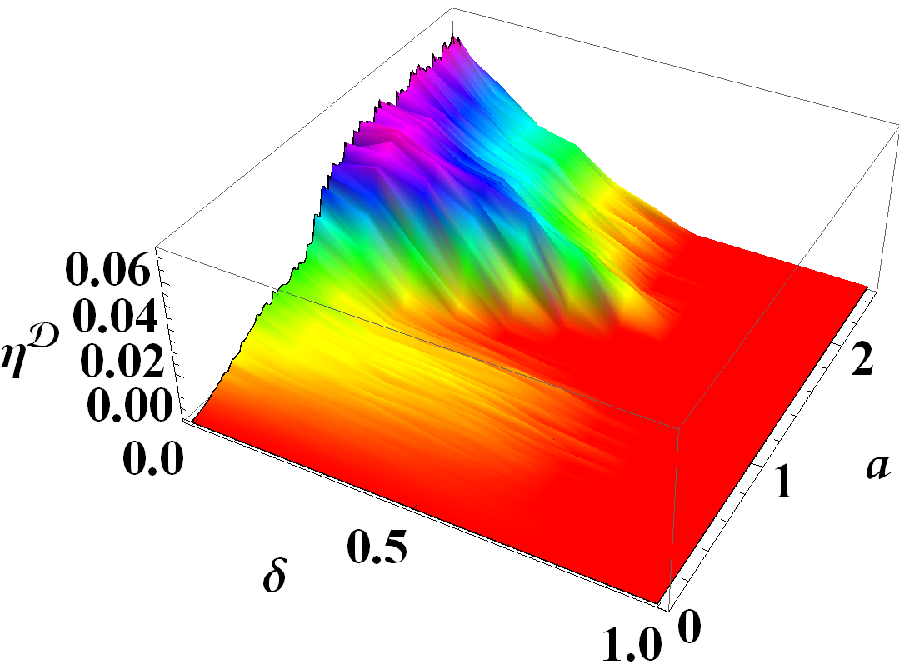}
 \caption{(Color online) Ergodicity score for quantum discord.  The ergodicity score for quantum discord of the anisotropic Heisenberg XYZ chain (with a magnetic field) of $8$ spins, arranged in a ring, is plotted against $\delta$ and the applied initial magnetic field $a$, for a fixed $\gamma =0.8$. The initial state of the time-evolution is the $t=0$ canonical equilibrium state at a temperature given by $J\alpha=20$. The ergodicity score is measured in bits. All other physical parameters used in the figure are dimensionless.}
 \label{fig:ergoscoredis1d}
\end{figure}


We now examine the ergodicity properties of the quantum correlation measures. From Figs. \ref{fig:eqentgadel} and \ref{fig:evoldistimedelta}, by analyzing the behavior of the entanglements of the equilibrium and evolved states, we find that entanglement measures are ergodic for all values of $\delta$, $\gamma(\neq 0)$, and $a$. We have analyzed this for logarithmic negativity as well as for concurrence. Hence, the ergodicity score is vanishing for all system parameters for all such measures.

Quantum discord and quantum work-deficit, both information-theoretic measures, also remain ergodic, when $\delta \geq \gamma$. However, for $\delta < \gamma$ these measures exhibit nonergodicity for a large range of the magnetic field. In Fig. \ref{fig:ergoscoredis1d}, we plot $\eta^{\cal D}$ with respect to the $\delta$ and the field strength, $a$, for $\gamma=0.8$, where we assume that the time-evolution starts off from the canonical equilibrium state for the Hamiltonian in Eq. (\ref{Hamilwithfield}) at $t=0$ and for temperature given by $J\alpha=20$. To plot $\eta^{\cal D}$, we choose $J\beta=20$ for the equilibrium state, in the calculation of ${\cal Q}^{\scriptsize{can}}(T, h(t=\infty))$, since we find that the quantum discord of the equilibrium state is a monotonically increasing function with respect to $J\beta$ and saturates for a $J\beta$ much below $J\beta=20$. 

\begin{figure}[h!]
 \includegraphics[angle=-90,width=0.238\textwidth,keepaspectratio]{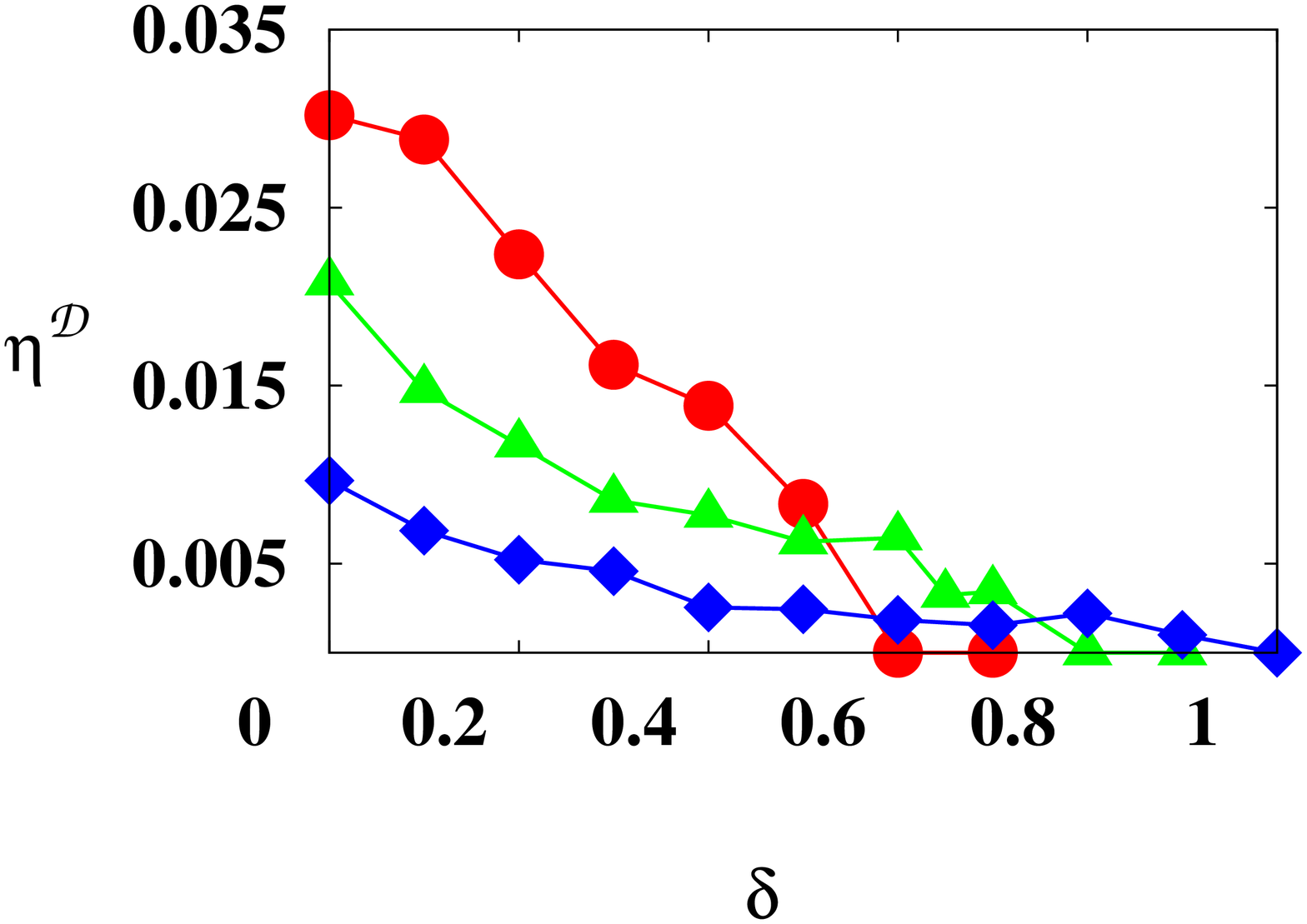}
 \includegraphics[angle=-90,width=0.238\textwidth,keepaspectratio]{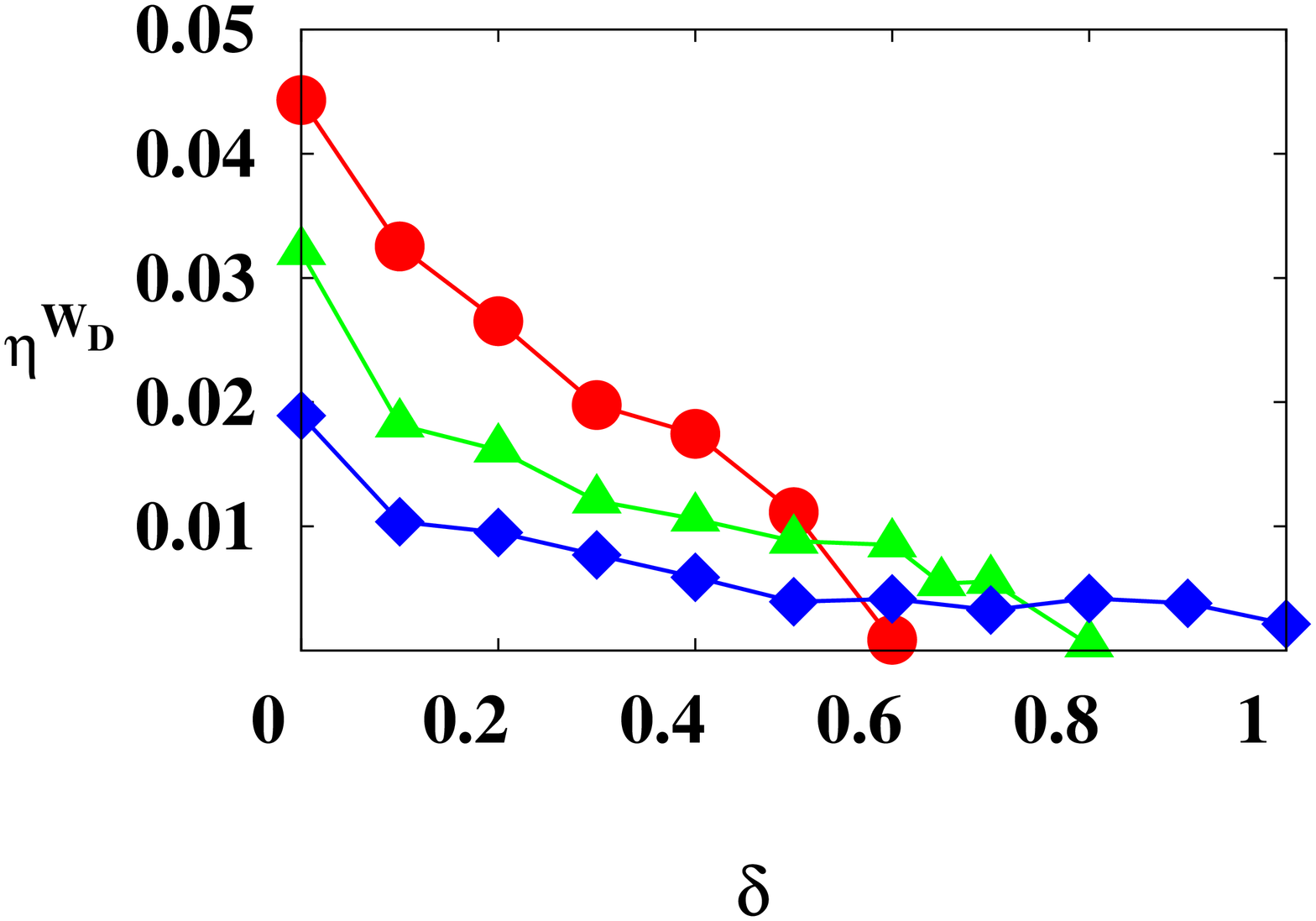}
 \caption{(Color online). Comparing ergodicity scores for quantum discord and quantum work-deficit.
 The ergodicity score for quantum discord (left) and quantum work-deficit (right) of the nearest-neighbor reduced state of the time-evolved states of the anisotropic Heisenberg XYZ chain (with a magnetic field) of 12 spins, arranged in a ring, is plotted against $\delta$, for different values of $\gamma$ and a for fixed initial magnetic field $a=0.6$. Here we choose $J\alpha =20$ for the $t=0$ canonical equilibrium state from which the evolution starts off. 
The depicted curves are for $\gamma =0.4$ (red circles), $\gamma =0.6$ (pink triangles), and $\gamma =0.8 $ (green squares). 
The ergodicity score for quantum discord is measured in bits, while that for quantum work-deficit is measured in qubits. All other quantities used in the figure are dimensionless.
} 
\label{fig:diswddelh}
\end{figure}

The trends, with respect to $\delta$, of ergodicity scores of quantum discord and quantum work-deficit for different $\gamma$, are depicted in Fig. \ref{fig:diswddelh}. For a fixed anisotropy $\gamma$, there always exists a certain value of $\delta$, for which quantum discord changes from being nonergodic to being ergodic. We denote that critical value of $\delta$ as $\delta_c^{\gamma}$, remembering that it pertains to quantum discord, and that there is a similar critical $\delta$, at a possible different value, for quantum work-deficit. We observe that the $\delta_c^{\gamma}$ increases with the increase in $\gamma$, and in Fig. \ref{fig:diswddelh}, $\delta_{c}^{\gamma=0.4} < \delta_{c}^{\gamma=0.6} < \delta_{c}^{\gamma=0.8} $ for both quantum discord and quantum work-deficit. 

The general behavior, of the quantum correlation measures in this system, that is emerging, is as follows. Entanglement measures exhibit ergodic behavior in all relevant parameter domains. The picture is richer for information-theoretic quantum correlation measures, and in particular, for a given anisotropy $\gamma$ and a given measure, there is a critical $\delta=\delta_c^{\gamma}$ at which the system transits from nonergodic to ergodic behavior for that measure.

\section{Quantum Heisenberg XYZ spin ladder with magnetic field }
 
It is interesting to study whether the two quantum correlation paradigms showing opposing statistical mechanical behavior 
persists in  higher-dimensional systems. To find this, we first consider the spins in a ladder arrangement, which is made up of two Heisenberg XYZ spin-$\frac{1}{2}$ chains, coupled by the same interactions along the rungs \cite{DagottoRice}. There is the time-dependent $z$-field at all sites. Periodic boundary condition is assumed along the rails. Such systems can be found in solid state materials  like \(\mbox{Sr}_{2}\mbox{CuO}_{3}\) and \(\mbox{Sr}_{14}\mbox{Cu}_{24}\mbox{O}_{41}\) \cite{Lemmensa03}.
 Recently it was found that the entanglement spectrum \cite{Calabrese2008} of the ground state of this model is related to the energy spectrum of its two single Heisenberg chain \cite{Poilblanc}.

\begin{figure*}%
\includegraphics[width=0.35\textwidth]{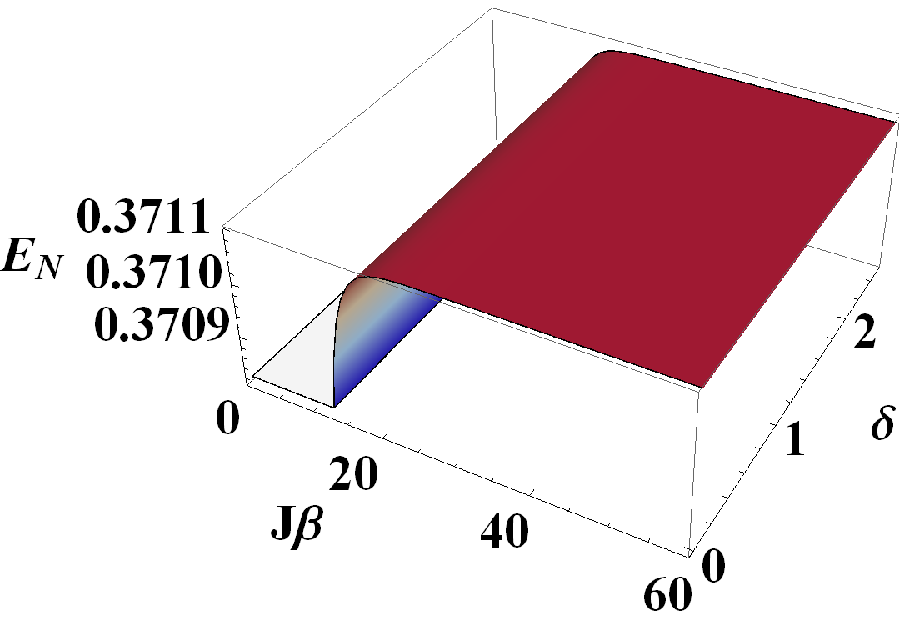}%
\includegraphics[width=0.35\textwidth]{./plots/entvsbetadeltaLad0o2}\\%
\includegraphics[width=0.35\textwidth]{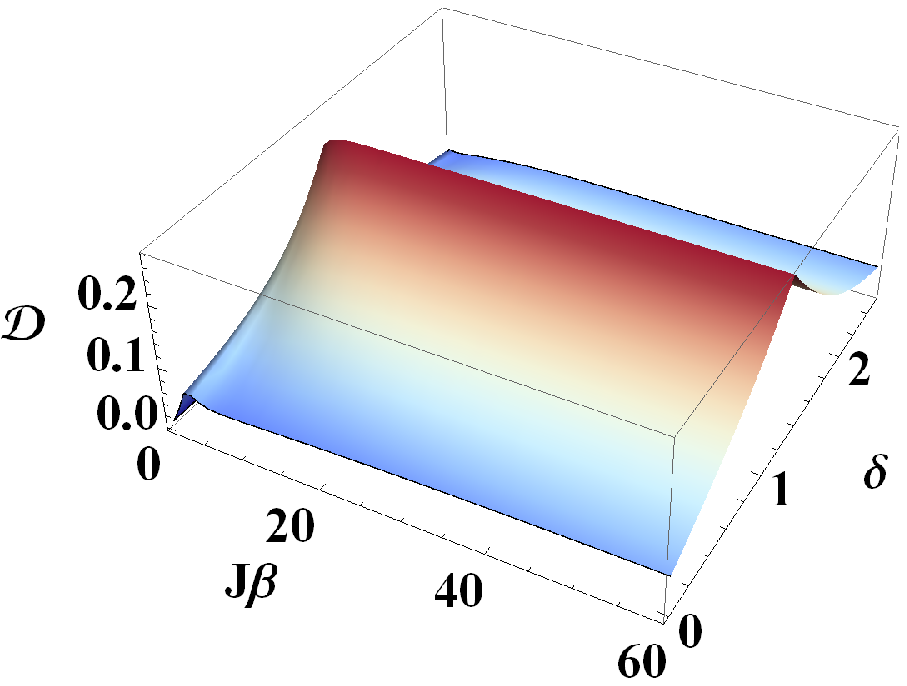}%
\includegraphics[width=0.35\textwidth]{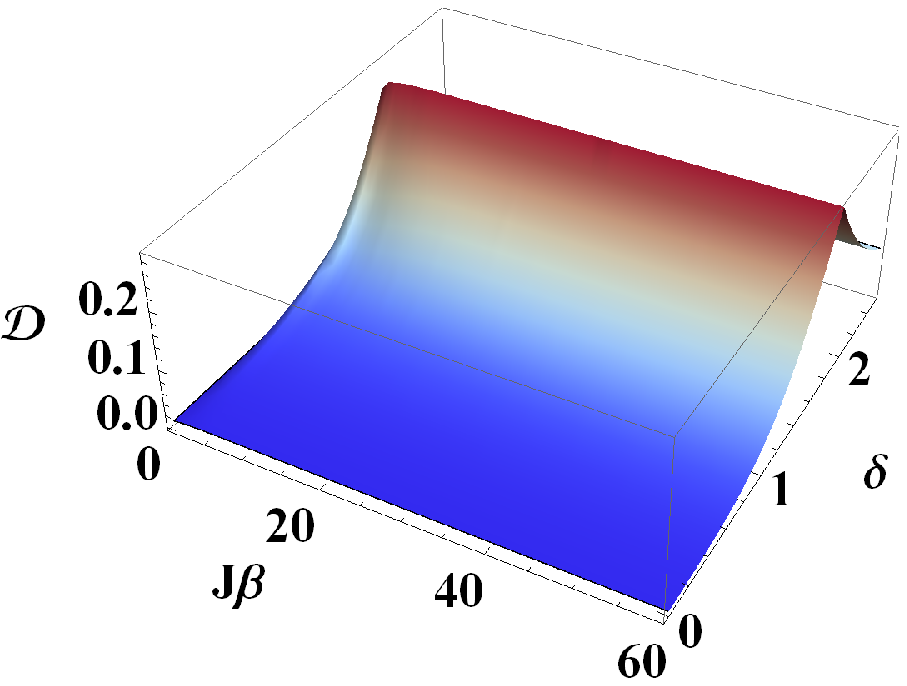}%
\caption{(Color online)  Behavior of quantum correlations in the equilibrium state. We plot quantum correlation measures of nearest-neighbor reduced states of the canonical equilibrium states, for a system of 8 quantum spin-$\frac{1}{2}$ particles arranged as a ladder and described by the Hamiltonian $H$ with respect to  $J\beta$, and the relative strength of the $zz$-interaction, \(\delta\), for different values of \(\gamma\). The top plots are for logarithmic negativity and the bottom ones are for quantum discord. The left plots  are for \(\gamma=0.2\) while the right one are for \(\gamma=0.8\). Quantum discord is measured in bits. All other axes in the figures correspond to dimensionless parameters.}%
\label{ladbetadelta}%
\end{figure*}

\begin{figure}
 \includegraphics[angle=-90,width=0.238\textwidth,keepaspectratio]{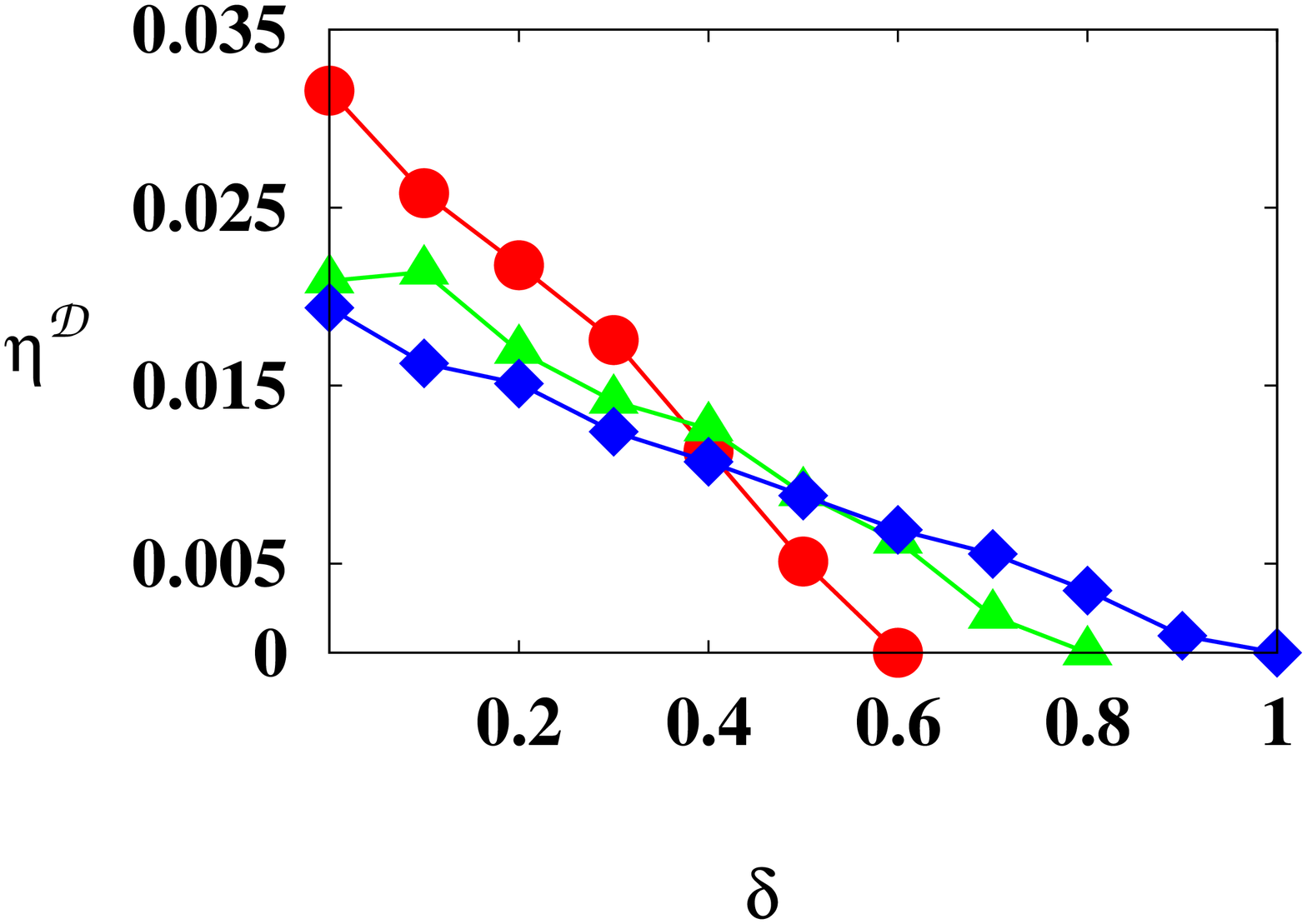}
 \includegraphics[angle=-90,width=0.238\textwidth,keepaspectratio]{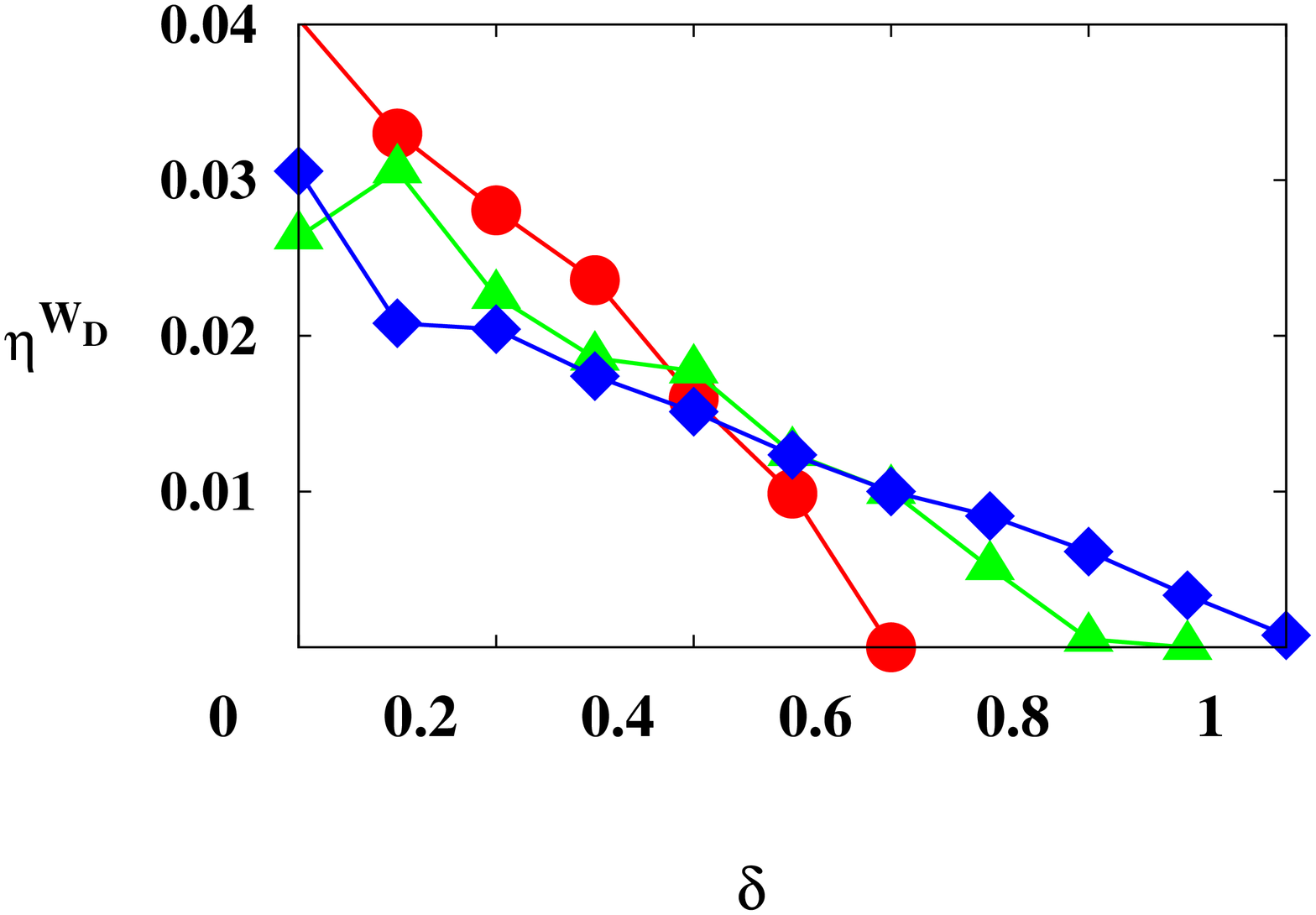}
 \caption{(Color online). Ergodicity curves in the Heisenberg XYZ ladder. The ergodicity scores of quantum discord (left) and quantum work-deficit (right) of a nearest-neighbor reduced state, along a rail, of the time-evolved state, in the ladder, of 8 spins is plotted with respect to the relative strength of the $zz$-interactions. The transition points, where the system moves from nonergodic to ergodic behavior of the information-theoretic measures are qualitatively similar to those in one-dimension, for a fixed $\gamma$. The depicted plots are for $\gamma =0.4$ (red circles), $\gamma =0.6$ (pink triangles) and $\gamma =0.8 $ (green squares). For the evolved state, $a=0.6,\, J\alpha=20$. The units are the same as in Fig. \ref{fig:diswddelh}.}
 \label{fig:ladder}
\end{figure}

In this model, the quantum correlation  measures of the evolved  and equilibrium states behave in a similar fashion as for the XYZ chain. In particular, entanglement of the nearest-neighbor states remain ergodic in this case. And there exists a critical $\delta$, above which the time-averaged value of the information-theoretic correlation measures, quantum discord and quantum work-deficit, of the nearest-neighbor reduced states of the evolved states match with 
the same measure of the equilibrium state, for some $\beta$, in a given magnetic field and a given $\gamma$ (see Fig. \ref{fig:ladder}  for the states along the rails). Quantum discord of the long-time equilibrium state does not remain a monotonically increasing function with $\beta$ like in the $1D$ model. 
See Fig. \ref{ladbetadelta}. To calculate $\eta^{\cal D}$,  we choose $J\beta=60$, at which the maximum value of ${\cal Q}^{\scriptsize{can}}(T, h(t=\infty))$ is attained,  for all values of $\delta$.
$\delta_c^{\gamma}$ increases with the increase in $\gamma$, while it is independent of the choice of the initial applied magnetic field for a fixed $\gamma$. These qualitative features of quantum discord remain the same, when a rung of the ladder is considered.  
A similar feature is observed for quantum work-deficit of the rung and rail states. See Fig. \ref{fig:ladder}(b) in this respect. We therefore again find that the strength of the $zz$-interaction, as quantified by $\delta$, can be adjusted in such a way that
the nonergodic nature of the information-theoretic measures, that persists in this system for low $\delta$, gets washed off, and we obtain ergodic behavior for high $\delta$.

\section{2D Quantum Heisenberg XYZ model with magnetic field}

\begin{figure}
 \includegraphics[angle=-90,width=0.238\textwidth,keepaspectratio]{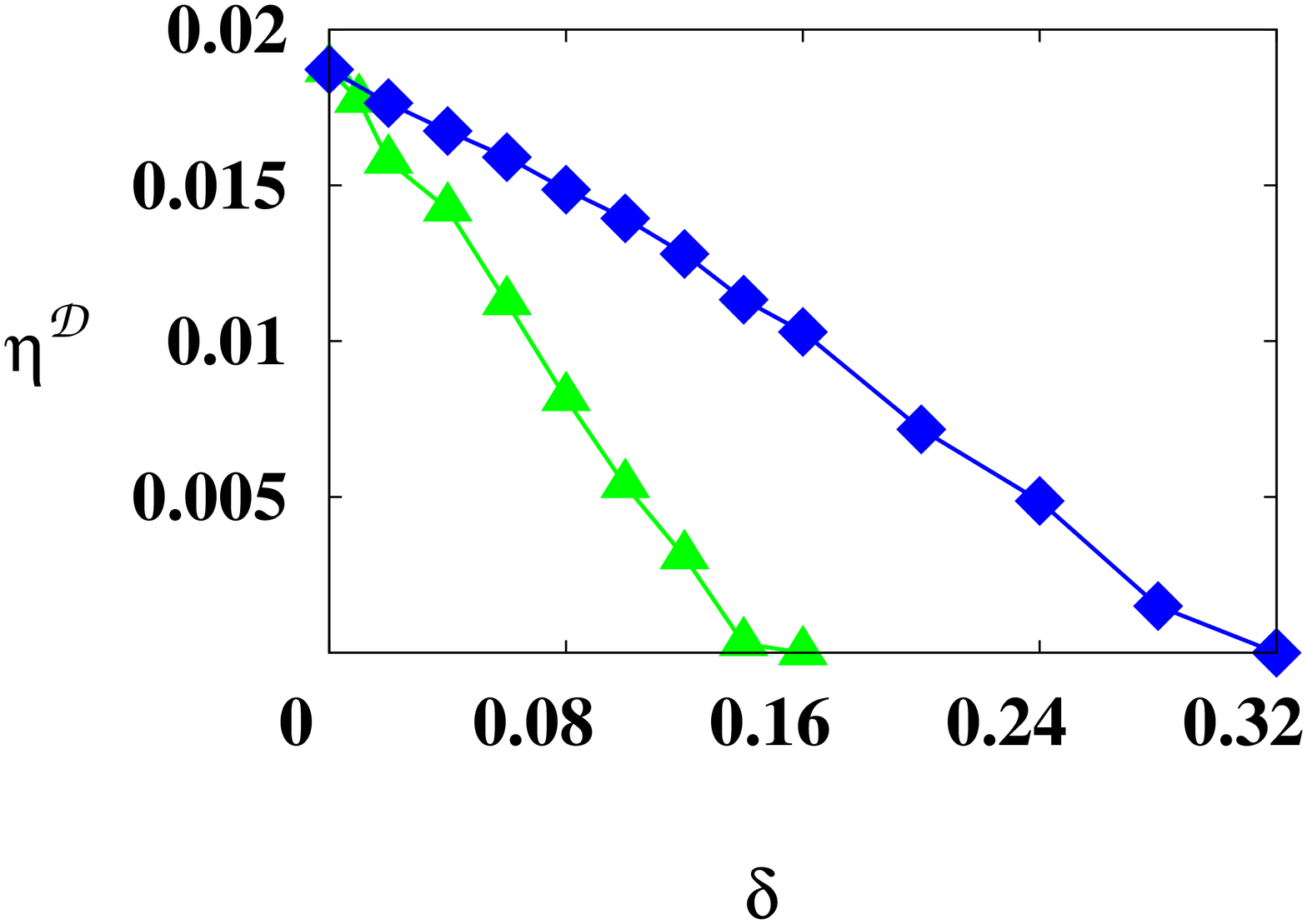}
 \includegraphics[angle=-90,width=0.238\textwidth,keepaspectratio]{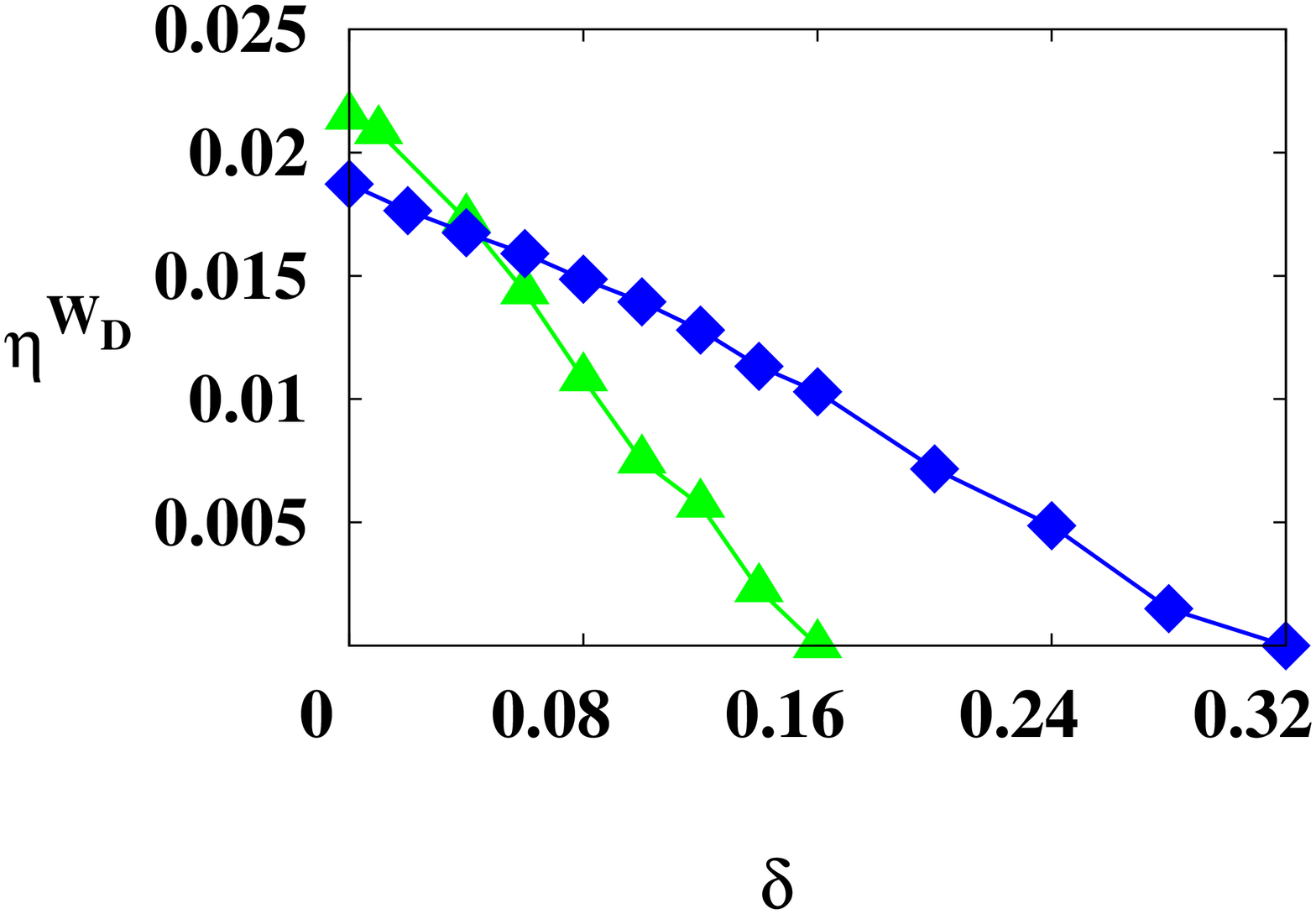}
 \caption{(Color online). Ergodicity scores in the $2D$ Heisenberg XYZ model. The ergodicity scores of quantum discord (left) and quantum work-deficit (right) in the nearest-neighbor reduced state of the time-evolved state, with respect to the strength of the $zz$-interaction for the anisotropic  Heisenberg XYZ model on a $2D$ square lattice, consisting of $12$ spins in a torus. The plots are for $\gamma =0.6$ (pink triangles) and $\gamma=0.8 $ (green squares). For the time-evolved state, $a=0.6,\, J\alpha=20$. The units are the same as in Fig. \ref{fig:diswddelh}.}
\label{fig:2d}
\end{figure}

The two-dimensional Heisenberg model describes important systems, including materials like \(\mbox{SrCu}_{2}(\mbox{BO}_{3})_{2}\) and
 \(\mbox{CaV}_{4}\mbox{O}_{9}\) \cite{Miyahara}.
Experimental studies of the Heisenberg model in $2D$ lattices have been proposed e.g., in trapped ions \cite{H2dion} and optical lattices \cite{H2doptlat}.

We consider a quantum Heisenberg XYZ spin model on a square lattice with antiferromagnetic  interactions between the nearest-neighbor spins. Periodic boundary condition is assumed and hence, geometrically, the system forms a spin-arrangement on a torus. The time-dependent magnetic field is assumed to be active at all sites.  Like in the ladder and $1D$ models, we again find that the entanglement measures are ergodic for all values of $\gamma$, $\delta$, and the initial magnetic field $a$. Interestingly, unlike in the $1D$ and ladder systems, the transition from nonergodicity to ergodicity of the information-theoretic measures, occurs for relatively low values of the $zz$-interaction strength, i.e., for low values of $\delta$ (Fig. \ref{fig:2d}). For example, when $\gamma \,\, \mbox{and}\,\, h$ are  0.6, in the ladder and $1D$ systems, both quantum discord and quantum work-deficit remain nonergodic till $\delta \approx 0.8$, while they both become ergodic in $2D$ at $\delta \approx 0.16$.
These observations lead us to infer that information-theoretic measures are more sensitive to the dimension of the lattice,  than the entanglement measures, with respect to their statistical mechanical properties.

\section{Discussion}

Quantum Heisenberg models have created lot of interest due to their rich physical properties and the possibility of realizing such systems in artificial materials as well as in inorganic compounds. However, investigations into the dynamics of such models, for example, under the influence of time-dependent magnetic fields,  are limited by the fact that the system cannot be diagonalized analytically. Here, we have studied the behavior of quantum correlations, both from the entanglement-separability paradigm and the information-theoretic one, of the equilibrium state as well as the evolved state of the quantum Heisenberg anisotropic XYZ model, by numerical simulations. In particular, we found that although entanglement measures are ergodic irrespective of the system parameters, information-theoretic measures exhibit a rich picture, with respect to their statistical mechanical properties. Specifically, we find that the $zz$-interaction strength has a cross-over value, for a given $xy$-anisotropy and a given information-theoretic quantum correlation measure, that indicates a transition from nonergodic to ergodic behavior for that measure. The qualitative features of the measures in the entanglement-separability paradigm and the information-theoretic one are the same in the one-dimensional, ladder, and two-dimensional square lattices. However, in the square lattice, the information-theoretic measures are more sensitive to the change of the $zz$-interaction strength than in other dimensions. Such dimension-dependent change of ergodic behavior is absent for entanglement measures.

\acknowledgements
R.P. acknowledges support from the Department of Science and Technology, Government of India, in the form of an INSPIRE faculty scheme at the Harish-Chandra Research Institute (HRI), India. We acknowledge computations performed at the cluster computing facility in HRI.

\end{document}